\documentstyle[epsf,psfig,12pt]{article}

\def\spose#1{\hbox to 0pt{#1\hss}}
\def\lsim{\mathrel{\spose{\lower 3pt\hbox{$\mathchar"218$}}
 \raise 2.0pt\hbox{$\mathchar"13C$}}}
\def\gsim{\mathrel{\spose{\lower 3pt\hbox{$\mathchar"218$}}
 \raise 2.0pt\hbox{$\mathchar"13E$}}}

\begin{document}
\setlength{\baselineskip}{23pt}

\begin{center}
\Large\bf{Isospin Symmetry Breaking
in the Quark Condensates,  
$\langle \bar dd\rangle\not= \langle \bar uu\rangle$, as Estimated
from QCD Sum Rules }
\end{center}

\vspace{0.5cm}
\begin{center}
W-Y. P. Hwang$^\dagger$ and Kwei-Chou Yang$^\ddagger$\\
{\sl $^\dagger$Department of Physics, National Taiwan University, Taipei,
Taiwan 10764, R.O.C.\\
$^\ddagger$Institute of Physics, Academia Sinica, Taipei, Taiwan 115, R.O.C.}
\end{center}


\vspace{0.7cm}
\begin{abstract}
\vspace{0.2cm}\noindent
Isospin symmetry breaking in the quark condensates, $\langle \bar dd\rangle 
\not= \langle \bar uu\rangle$, is a fundamental parameter in both the QCD sum 
rule studies and the chiral perturbation theory. In this article, we 
apply the QCD sum rule method to treat the hyperon mass difference
${1\over 2}(m_{\Sigma^-}+m_{\Sigma^0})-m_{\Sigma^+}$ in order to obtain a
reasonable estimate on such isospin symmetry breaking parameter $\gamma$ 
defined as $(\langle \bar dd\rangle-\langle \bar uu\rangle)/ 
\langle \bar uu\rangle$. Note that the electromagnetic contributions to the
particular mass difference are expected to cancel almost completely. Using 
the light-quark mass difference $m_d-m_u=4\pm 1$ MeV and the experimental data
${1\over 2}(m_{\Sigma^-}+m_{\Sigma^0})-m_{\Sigma^+}$$=5.62\pm 0.13$ MeV,
we obtain $\gamma= -0.011\pm 0.001$, a value which is slightly larger than the 
commonly adopted value and is of smaller uncertainty.
\end{abstract}
\vskip 2cm
{PACS number(s): 11.30.Hv; 12.38.Lg; 11.55Hx; 14.20.-c}


\newpage
\setlength{\baselineskip}{23pt}
Although quantum chromodynamics (QCD) has been accepted as a theory for
describing strong interactions among quarks and gluons, 
it is highly nonperturbative at the confinement scale. In particular, it is
expected that the QCD ground state, or the QCD vacuum, has nonzero 
quark condensate $<{\bar q} q>$, gluon condensate $<G^2>$, and many other
condensates of higher order. Making use of the method  
QCD sum rules\cite{svz}, it has become possible to study 
hadron physics in terms of the basic parameters of QCD, such as current 
quark masses and non-perturbative condensates. In a systematic study of isospin 
symmetry breaking effects, both the quark mass difference, $m_d-m_u$, and
the difference in condensates, $\gamma \equiv (\langle \bar dd\rangle/\langle 
\bar uu\rangle)-1$ are the most fundamental parameters. while the quark mass
difference $m_d-m_u$ can be estimated with reasonable confidence as
$m_d-m_u\approx 3-5$ MeV\cite{lw,gl}, $\gamma$ is poorly known.
Theoretical estimates on $\gamma$ have been obtained in the chiral 
perturbation theory\cite{gl2}, in the Nambu-Jona-Lasinio model\cite{hhp,ab}, 
and in QCD sum rules\cite{cz,yh1,adi,ei,fn}; the value has a fairly large 
range, $\gamma=-(2\sim 10)\times 10^{-3}$. In the QCD sum rule approach,
the error in estimating the value of $\gamma$ comes from the poorly-known
electromagnetic contributions\cite{yh1} and instanton effects\cite{fn}.

In this paper, we wish to use the QCD sum rule method to obtain the isospin 
symmetry breaking parameter $\gamma$ from the hyperon mass difference
${1\over 2}(m_{\Sigma^-}+m_{\Sigma^0})-m_{\Sigma^+}$, a combination that
the electromagnetic contributions are expected to cancel almost exactly. The 
cancelation may be understood as follows. In the simple (constituent)
quark model \cite{mi,dgg,lw,gl}, the electromagnetic contributions can be 
separated into a contribution from photon exchange between
two of the three quarks and a contribution from the quark
self-energy type:
\begin{eqnarray}
m^\gamma =\alpha \langle 1/r\rangle (Q_1 Q_2+Q_2 Q_3+Q_3 Q_1)+
C(Q_1^2+Q_2^2+Q_3^2),
\end{eqnarray}
where $Q_1$, $Q_2$, and $Q_3$ are the quark charges and $\langle 1/r\rangle$
is the (effective) average inverse distance of two quarks in a baryon.
In the isospin symmetric limit, the radii of $\Sigma^+$, $\Sigma^-$, and 
$\Sigma^0$ are identical so that 
\begin{eqnarray}
&&\langle 1/r\rangle_{\Sigma^+}\approx\langle 1/r\rangle_{\Sigma^-}
\approx\langle 1/r\rangle_{\Sigma^0},\nonumber\\
&& C^{\Sigma^+}\approx C^{\Sigma^-}\approx C^{\Sigma^0}.
\end{eqnarray}
Here the equality would take hold should isospin symmetry be exact.
We therefore obtain
\begin{eqnarray}
\frac{1}{2}(m^\gamma_{\Sigma^-}+m^\gamma_{\Sigma^0})-m^\gamma_{\Sigma^+}=0.
\end{eqnarray}
We note that similar arguments may be constructed for other models of hadrons.
Using the quark model estimates and the experimental information
from electron scattering on the protons and neutrons (extrapolated to the
other members of the baryon octet), Gasser and Leutwyler\cite{gl} have reported
$\frac{1}{2}(m^\gamma_{\Sigma^-}+m^\gamma_{\Sigma^0})-m^\gamma_{\Sigma^+}=
-0.32\pm 0.26$ MeV, which is consistent with the above observation. 
Accordingly, the experimentally observed value, ${1\over 2}(m_{\Sigma^-}
+m_{\Sigma^0})-m_{\Sigma^+}$$ =5.62\pm 0.13$ MeV, may serve to
determine $m_d-m_u$ and $\gamma$ in an unambiguous manner. 
As some authors have indicated that 
instanton contributions are absent in the chirally even baryon sum 
rules\cite{fn} but may be of significance in the chirally odd one, we choose 
to employ solely the chirally even sum rules.

We turn our attention to the derivation of QCD sum rules. Following the 
standard procedure, we consider the two-point Green's function
\begin{eqnarray}
\Pi(p)&=i \int d^4x\ e^{ipx}
\bigl<0\vert T\bigl(\eta(x) \bar\eta(0)
\bigl)\vert 0\bigl>,
\end{eqnarray}
where the composite field operators are specified by
\begin{eqnarray}
\eta_{\Sigma^+}=&&\epsilon^{abc}({u^a}^TC\gamma_\mu
  u^b)\gamma_5\gamma^\mu s^c,\nonumber\\
\eta_{\Sigma^0}=&&\epsilon^{abc}{1\over\sqrt2}[({u^a}^TC\gamma_\mu
  d^b)\gamma_5\gamma^\mu s^c+({d^a}^TC\gamma_\mu
  u^b)\gamma_5\gamma^\mu s^c],\nonumber\\
\eta_{\Sigma^-}=&&\epsilon^{abc}({d^a}^TC\gamma_\mu
  d^b)\gamma_5\gamma^\mu s^c,
\end{eqnarray}
with $a$, $b$, and $c$ color indices and $C$=$-C^T$ the charge conjugate
operator.
At the hadronic level, we write the Green's  function via dispersion relation as
\begin{eqnarray}
\Pi(p)
={1\over \pi}\int_0^\infty{\hbox{Im}\ \Pi(p')
\over p'^2-p^2-i\varepsilon}dp'^2,
\end{eqnarray}
where
\begin{eqnarray}
\hbox{Im}\ \Pi(p')=\pi \sum_X \delta(p'^2-M_X^2)
  \bigl<0\vert\eta(0)\vert X\bigl>\bigl<X
  \vert\bar\eta(0)
  \vert 0\bigl>,
\end{eqnarray}
and
\begin{eqnarray}
\langle 0|\eta(0)|\Sigma\rangle =\lambda_\Sigma v(p),
\end{eqnarray}
with $v(p)$ the Dirac spinor normalized as $\bar v(p) v(p)$=2$m_\Sigma$.
At the quark level, we write, using the $\Sigma^+$ case as our example,
\begin{eqnarray}
\bigl<0|T(\eta(x) \bar\eta(0))|0\bigl>
=2i\epsilon^{abc}\epsilon^{a'b'c'}Tr[S^{aa'}_u(x)\gamma_\nu C
(S^{bb'}_u(x))^T C\gamma_\mu]\gamma_5\gamma^\mu S^{cc'}_d(x)\gamma^\nu\gamma_5.
\end{eqnarray}
Here we have used the definition,
$$iS^{ab}(x)\equiv \bigl<0|T[q^a(x) \bar q^b(0)]|0\bigl>,$$
as in Ref. \cite{yh1}.

$\Pi(p)$ has two invariant structures:
\begin{eqnarray}
\Pi(p)=\Pi_1(p^2)+\Pi_{\not p}(p^2)\not p,
\end{eqnarray}
where $\Pi_1$ and $\Pi_{\not p}$ are two invariant functions. We name
$\Pi_1$ as the chirally odd sum rule since chiral-odd operators 
contribute dominantly, and $\Pi_{\not p}$ as the chirally even sum rule.
The derivation of the relevant QCD sum rules is similar to what we have 
reported in our earlier papers \cite{yh1,yh2}. Here we record only the final 
results on the $\Pi_{\not p}$ (chirally even) $\Sigma$ mass sum rules:
\begin{eqnarray}
\beta_{\Sigma ^+}^2 e^{-(M_{\Sigma ^+}^2/ M^2)}=&&
{M^6\over 8}L^{-{4/9}}E_2^{\Sigma^+} +{bM^2\over 32}L^{-{4/9}}E_0^{\Sigma^+}
   +{a_u^2\over 6}L^{4/9}-{a_u^2 m_0^2\over 24M^2}L^{-{2/27}}\cr
 &&-{m_s a_s M^2\over 4}L^{-{4/9}}E_0^{\Sigma^+}
  -{m_s a_s m_0^{\prime 2}\over 24}L^{-{26/27}}
 -{m_u a_u m_0^2\over 12}L^{-{26/27}},
\end{eqnarray}
\begin{eqnarray}
\beta_{\Sigma ^0}^2 e^{-(M_{\Sigma ^0}^2/ M^2)}=&&
{M^6\over 8}L^{-{4/9}}E_2^{\Sigma^0}+{bM^2\over 32}L^{-{4/9}}E_0^{\Sigma^0}
   +{a_u a_d\over 6}L^{4/9}-{a_u a_d m_0^2\over 24M^2}L^{-{2/27}}\cr
 &&-{[m_s a_s-(a_u-a_d)(m_d-m_u)]M^2\over 4}L^{-{4/9}}E_0^{\Sigma^0}
  -{m_s a_s m_0^{\prime 2}\over 24}L^{-{26/27}}\cr
 &&-{[a_d(3m_u-m_d)+a_u(3m_d-m_u)]m_0^2\over 48}L^{-{26/27}},
\end{eqnarray}
and
\begin{eqnarray}
\beta_{\Sigma ^-}^2 e^{-(M_{\Sigma ^-}^2/ M^2)}=&&
{M^6\over 8}L^{-{4/9}}E_2^{\Sigma^-}
          +{bM^2\over 32}L^{-{4/9}}E_0^{\Sigma^-}
   +{a_d^2\over 6}L^{4/9}-{a_d^2 m_0^2\over 24M^2}L^{-{2/27}}\cr
 &&-{m_s a_s M^2\over 4}L^{-{4/9}}E_0^{\Sigma^-}
  -{m_s a_s m_0^{\prime 2}\over 24}L^{-{26/27}}
 -{m_d a_d m_0^2\over 12}L^{-{26/27}},
\end{eqnarray}
where we have adopted the definitions:
\begin{eqnarray}
a_q&&\equiv -{(2\pi)}^2\bigl<\bar qq\bigl>,\nonumber\\
a_qm_0^2&&\equiv{(2\pi)}^2\bigl<g_s\bar q\sigma\cdot Gq\bigl>,\quad
q=u\ \hbox{\rm or}\ d,\cr
 b&&\equiv \langle g_s^2G^2\rangle,\cr
            a_s&&\equiv -{(2\pi)}^2\langle\bar ss\rangle,\nonumber\\
a_s m_0'^2&&\equiv(2\pi)^2\langle g_s\bar s\sigma\cdot Gs\rangle,\cr
\beta^2_\Sigma&&\equiv {(2\pi)}^4{\lambda^2_\Sigma\over 4},\cr
        E_n^{(m)}&&\equiv 1-e^{-x_{(m)}}(1+x_{(m)}+\dots+{1\over n!}x_{(m)}^n)
                 ,\quad\hbox{\rm with}\ x_{(m)}=W_{(m)}^2/M^2,\nonumber\\
L&&\equiv {{\rm ln}(M^2/\Lambda^2)\over {\rm ln}(\mu^2/\Lambda^2)}.
\end{eqnarray}
In Eqs. (11-13), $M^2$ is the Borel mass squared, as arising from the Borel
transform of the sum rules,
\begin{eqnarray}
B[\Pi_{1(\not p)}(p^2)] \equiv \lim_{{\scriptstyle n\to\infty \atop\scriptstyle -p^2\to\infty }
\atop\scriptstyle {-p^2\over nM^2}\ fixed}\frac{1}{n!}(-p^2)^{n+1}
 \bigl({d\over dp^2}\bigl)^n \Pi_{1(\not p)}(p^2).
\end{eqnarray}
In what follows, we take the value $\langle\bar ss \rangle$
=0.8 $\times (\langle\bar uu\rangle+\langle\bar dd \rangle)/2$.
We also use the relation
$$\langle g_s\bar s\sigma \cdot Gs\rangle /\langle g_s\bar u\sigma\cdot Gu\rangle
\simeq \langle\bar ss\rangle/ \langle\bar uu\rangle,$$
which indicates $m_0'^2\simeq m_0^2$, a relation which has been examined in 
some detail in octet baryons \cite{cz}.
The input parameters in the numerical analysis are taken to be
$\bar a_q$=$(a_u+a_d)/2$=0.545 GeV$^3$, $a_s$=0.8 $\bar a_q$, $b$=0.474 GeV$^4$,
$m_0^2$=0.6-0.8 GeV$^2$, $\bar m_q$=$(m_u+m_d)/2$= 7 MeV, $m_d-m_u$=3-5 MeV, 
and $m_s$=150-200 MeV. These values correspond to the QCD scale parameters 
$\Lambda$= 0.1 GeV, and the normalization point $\mu$= 0.5 GeV.
Applying the operator $M^4\partial \ell n/\partial M^2$ to both sides of 
Eqs. (11-13) (to get rid of the dependence on the parameter $\beta_\Sigma^2$), 
we deduce the QCD sum rule on the mass difference 
${1\over 2}(m_{\Sigma^-}+m_{\Sigma^0})-m_{\Sigma^+}$.
We fit the two sides of this sum rule by minimizing
$$\delta(M^2)=\sum_{\scriptstyle {M^2=1.0+0.005i} \atop\scriptstyle 
0\le i\le 60} |LHS-RHS|$$
within the Borel region 1.0 GeV$^2\le M^2 \le 1.3$ GeV$^2$, where the $LHS$ 
(the left-hand side of the sum rule) is the experimental value, while the $RHS$ 
(the right-hand side of the sum rule) is the operator-product-expansion results at the quark-gluon
level. In the numerical analysis, we adopt $W^{\Sigma^+}$=3.47 GeV$^2$\cite{yh2}
 and the experimental
data ${1\over 2}(m_{\Sigma^-}+m_{\Sigma^0})-m_{\Sigma^+}$=5.62$\pm$0.13 MeV.
Our results are 
$(W_{\Sigma^0}^2+W_{\Sigma^-}^2)/2$=3.483$\pm$0.002 GeV$^2$,
$W_{\Sigma^-}^2-W_{\Sigma^0}^2$=0.0030$\pm$0.0004 GeV$^2$, and
$\gamma=-0.011\pm0.001$. Theoretical uncertainties in our determination
of the isospin symmetry breaking parameter $\gamma$ come mainly from the 
uncertainty in the value of $m_d-m_u$ and
the experimental error in ${1\over 2}(m_{\Sigma^-}+m_{\Sigma^0})-m_{\Sigma^+}$.

In Fig. 1, we plot the $LHS$ and $RHS$ as a function of the Borel mass squared, 
$M^2$. The solid curve is the $LHS$ assuming the value  
${1\over 2}(m_{\Sigma^-}+m_{\Sigma^0})-m_{\Sigma^+}$=5.62 MeV.
The dotted curve is the $RHS$ with the optimized values, $\gamma=-0.0114$,
$W^2_{\Sigma^0}$=3.4816 GeV$^2$, and $W^2_{\Sigma^-}$=3.4846 GeV$^2$, 
where we have used, as the input, $m_0=0.8$ GeV$^2$, $m_s=175$ MeV, 
$m_d$=8.9 MeV, and $m_u$=5.1 MeV. 

Following the same procedure as outlined above and using the present value 
for $\gamma$, we have repeated the analysis of the sum rule for the 
neutron-proton mass difference within the Borel region 0.9 GeV$^2 \le M^2 \le$ 
1.1 GeV$^2$ \cite{yh1,yh2}. The result is illustrated in Fig. 2, in which the 
solid curve is the $LHS$ assuming the value 
$m_N^{QCD}-m_P^{QCD}$=2.693 MeV and the dotted curve is the $RHS$ with 
the optimized values, $\gamma=-0.0102$, $W^2_N$=2.2506 GeV$^2$ (neutron), 
$W^2_P$=2.25 GeV$^2$ (proton), $m_d$=8.9 MeV and $m_u$=5.1 MeV.
Such value implies that the electromagnetic 
contribution to the neutron-proton mass difference $m_N^\gamma-m_P^\gamma$   
is about $-1.3\sim-1.8$ MeV, a value in reasonable agreement with 
the naive quark model result \cite{gl}. Note that
our result on $\gamma$ is a little larger than that obtained
by Forkel and Nielsen\cite{fn}, who obtained 
$\gamma=-(0.8-1.0)\times 10^{-2}$ making use of 
$m_N^\gamma-m_P^\gamma$=$-0.76\pm$0.3 MeV\cite{gl} (corresponding to
i.e., $m_N^{QCD}-m_P^{QCD}=2.05\pm 0.3$ MeV) in the Borel region 
0.8 GeV$^2\le M^2\le$1.4 GeV$^2$.  Using their value on 
$m_N^\gamma-m_P^\gamma$ and adopting our approach together with the
same parameters as in the present paper, we obtain 
$\gamma=-(0.7-0.9)\times 10^{-2}$, a result consistent with that in
Ref.\cite{fn} and also with that in our earlier paper \cite{yh1}.
Nevertheless, the choice $m_N^{\gamma}-m_P^{\gamma}=-0.76\pm0.3$ MeV
can only be marginally consistent with the result obtained through the
analysis of the sum rule for 
${1\over 2}(m_{\Sigma^-}+m_{\Sigma^0})-m_{\Sigma^+}$, which gives a better
determination of $\gamma$ and leads to 
$m_N^\gamma-m_P^\gamma$ is about $-1.3 \sim -1.8$MeV.

In summary, we have presented a detailed analysis of
the isospin symmetry breaking parameter in the quark condensate,
$\gamma$, from the chirally even QCD sum rules on the mass difference
${1\over 2}(m_{\Sigma^-}+m_{\Sigma^0})-m_{\Sigma^+}$, for which
the instanton contributions are absent and
the electromagnetic contributions are expected to cancel almost exactly.
Our numerical analysis indicates that a value of $\gamma$ ($=-0.011 \pm 0.001$) 
(slightly larger in magnitude)
is preferred over the existing result, $\gamma = -(2 \sim 10) \times 10^{-3}$.
Using the present value of $\gamma$ and the observed neutron-proton mass
difference, we have also deduced the electromagnetic contribution to the 
neutron-proton mass difference as in the range $-1.3\sim -1.8$ MeV, in 
reasonable agreement with the quark model result.

\vspace{1cm}
This work was supported in part by the National Science
Council of R.O.C. under Grant No. NSC87-2112-M001-048 and
No. NSC87-2112-M002-031.

%
%

\newpage
\begin{figure}
\epsfbox{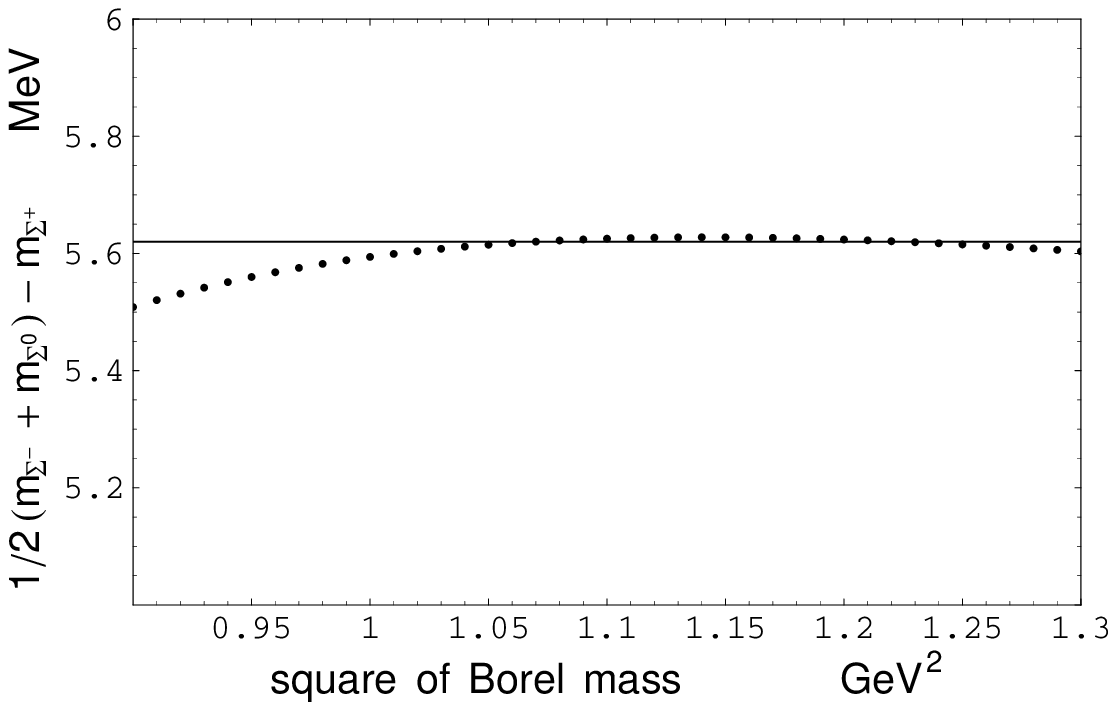}
\caption{\label{fig:trace1}
${1\over 2}(m_{\Sigma^-}+m_{\Sigma^0})-m_{\Sigma^+}$ as a function of 
the Borel mass squared $M^2$. The solid curve is the $LHS$, the left-hand side
of the sum rule, assuming 
${1\over 2}(m_{\Sigma^-}+m_{\Sigma^0})-m_{\Sigma^+}$=5.62 MeV.
The dotted curve is the $RHS$, obtained by making use of
$m_0=0.8$ GeV$^2$, $m_s=175$, $m_d$=8.9 MeV, and 
$m_u$=5.1 MeV as the input.}
\end{figure}
\vfill
\newpage
\begin{figure}
\epsfbox{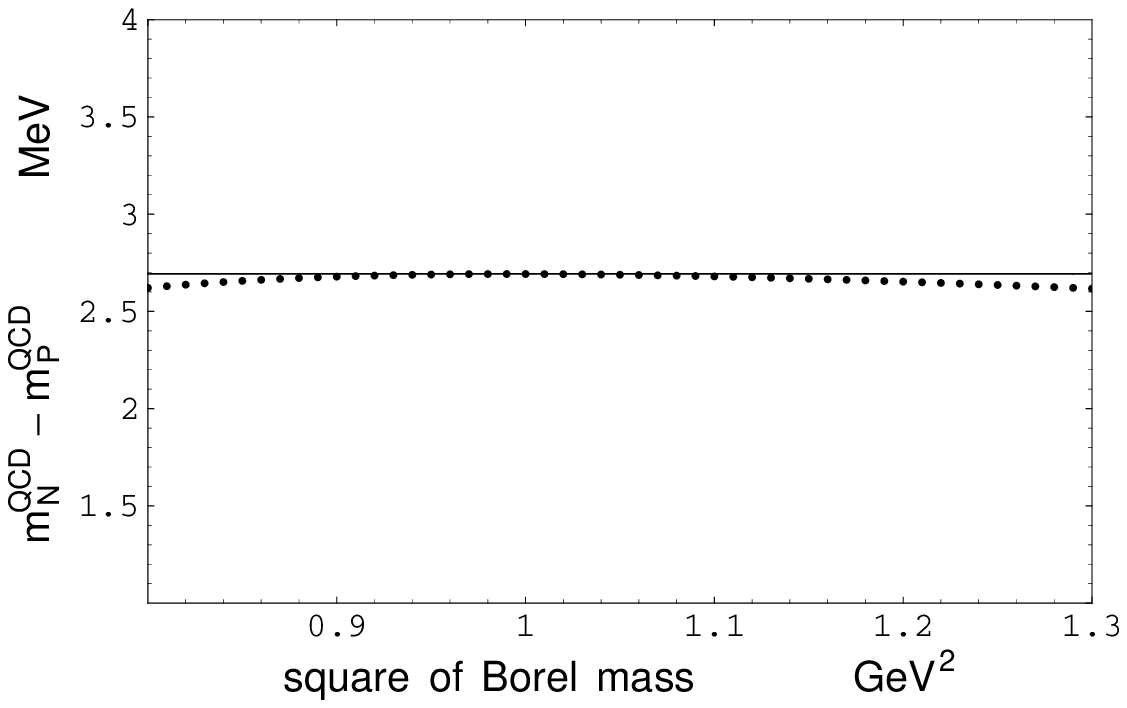}
\caption{\label{fig:trace2}
$m_N^{QCD}-m_P^{QCD}$ as a function of the Borel mass squared $M^2$.
The solid curve is the $LHS$ assuming
$m_N^{QCD}-m_P^{QCD}$=2.693 MeV and the dotted curve is the $RHS$ with 
the optimized values, $\gamma=-0.0102$ and $W^2_N$=2.2506 GeV$^2$. 
Here we have used $W^2_P$=2.25 GeV$^2$, $m_d$=8.9 MeV and $m_u$=5.1 MeV
as the input.} 
\end{figure}

\begin{thebibliography}{99}
\bibitem{svz} M. A. Shifman, A. I. Vainshtein, and V. I. Zakharov, Nucl. Phys.
B{\bf 147}, 385, 448 (1979); L. J. Reinders, H. Rubinstein, and S. Yazaki,
Phys. Rep. {\bf 127}, 1 (1985).

\bibitem{lw} K. Lane and S. Weinberg, Phys. Rev. Lett. {\bf 37}, 717 (1976).

\bibitem{gl} J. Gasser and H. Leutwyler, Phys. Rep. {\bf 87}, 77 (1982).

\bibitem{gl2} J. Gasser and H. Leutwyler, Nucl. Phys. {\bf B250}, 465 (1985).

\bibitem{hhp} T. Hatsuda, H. Hogaasen, and M. Prakash, Phys. Rev. C{\bf 42},
2212 (1990).

\bibitem{ab} C. Adami and G. E. Brown, Z. Phys. A {\bf 340}, 93 (1991).

\bibitem{cz} V. L. Chernyak and A. R. Zhitnitsky, Phys. Rep. {\bf 112},
173 (1984).

\bibitem{yh1} K.-C. Yang, W-Y. P. Hwang, E. M. Henley,
and L. S. Kisslinger, Phys. Rev. D{\bf 47}, 3001 (1993).

\bibitem{adi} C. Adami, E. G. Drukarev, and B. L. Ioffe, Phys. Rev. D{\bf 48},
2304, (1993).

\bibitem{ei} V. L. Eletsky and B. L. Ioffe, Phys. Rev. D{\bf 48}, 1441 (1993).

\bibitem{fn} H. Forkel and M. Nielsen, Phys.  Rev. D{\bf 55}, 71 (1997).

\bibitem{mi} Y. Miyamoto, Progr. Theor. Phys. {\bf 35}, 175 (1966).

\bibitem{dgg} A. De R$\acute{u}$jula, H. Georgi and S. L. Glashow, Phys. Rev.
D{\bf 12}, 147 (1975).

\bibitem{yh2} W-Y. P. Hwang and K.-C. Yang, Phys. Rev. D{\bf 49}, 460 (1994).

\end{thebibliography}
\end{document}